\documentclass[epi,twocolumn]{webofc}

\usepackage[varg]{txfonts}   
\wocname{EPJ Web of Conferences}
\woctitle{LHCP 2013}
\begin{document}
\title{Searching for supersymmetry in $Z'$ decays\thanks{Talk given at
LHCP 2013, Barcelona, Spain, May 13-18 2013.}}
%
%
\selectlanguage{english}

\author{Gennaro Corcella\inst{1}\fnsep\thanks{\email{gennaro.corcella@lnf.infn.it}}}

\institute{INFN, Laboratori Nazionali di Frascati,\\ Via E.~Fermi 40,
I-00044, Frascati (RM), Italy}

\abstract{I investigate production and decay of heavy neutral gauge bosons $Z'$ in GUT-inspired
U(1)$'$ groups and in the Sequential Standard Model. In particular, decays into
supersymmetric particles, such as slepton, chargino and neutralino pairs, as 
predicted in the MSSM, are accounted
for, with a special interest in final states with leptons and missing energy.
For a representative point of the parameter space, it is found that the inclusion
of supersymmetric decay modes has an impact of 200-300~GeV on the $Z'$ mass exclusion
limits.}
\maketitle
\section{Introduction}
\label{intro}
Searching for heavy neutral gauge bosons $Z'$ is one of the
main goals of the LHC experiments; in fact,
these bosons are particularly interesting since they 
are predicted by 
gauge models based on a U(1)$'$ symmetry, arising in the framework of
Grand Unification Theories (GUTs) \cite{rizzo,langacker}. 
The Sequential Standard Model (SSM), the simplest extension of the
Standard Model (SM), also contains extra heavy neutral bosons,
i.e. $Z'$ and $W'$, 
with the same couplings to fermions and bosons as the standard $Z$ and
$W$.
The ATLAS and CMS collaborations 
have searched for a $Z'$ at the LHC by looking at high-mass electron or muon 
pairs: in detail, with an integrated luminosity of 20 fb$^{-1}$,
CMS set limits $m_{Z'}>2.96$~TeV and $m_{Z'}>2.60$~TeV
on the mass of SSM and GUT-inherited $Z'$ bosons \cite{cms}, 
respectively, whereas 
the corresponding ATLAS limits are 2.86 TeV (SSM) and
in the range 2.38-2.54 TeV (U(1)$'$ models) \cite{atlas}.
However, such analyses completely rely on Standard Model $Z'$ decays:
allowing supersymmetric modes will substantially modify the
search strategies and the mass exclusion limits.

In fact, several reasons makes it compelling and interesting
investigating possible $Z'$ decays beyond the Standard Model (BSM).
Referring, for simplicity, to the Minimal Supersymmetric
Standard Model (MSSM), possible
$Z'$ decays into pairs of the lightest neutralinos
can lead to monojet or monophoton final states, which
are used to look for Dark Matter candidates.
From the point of view of supersymmetry, $Z'$ decays yield
a cleaner signal with respect to direct sparticle production in
$pp$ collisions, since, in events like $Z'\to \tilde\ell^+\tilde\ell^-,
\tilde\chi^+_i\tilde\chi^-_j,\dots$, $\tilde\ell^\pm$ and $\tilde\chi^\pm_{i,j}$
being sleptons and charginos, the $Z'$ mass gives a further
kinematic constrain on the invariant mass of the supersymmetric pair.
Furthermore, supersymmetric contributions to $Z'$ decays 
decrease the SM branching ratios and the number of expected high-mass
dimuons and dielectrons,
and therefore the exclusion limits on the $Z'$ mass, 
will have to be revisited. 
Ideally, if for a given scenario the BSM branching ratios
had to be comparable or even larger than the SM ones,
one could even argue that such decay channels may have hidden the
$Z'$ in the searches carried out so far.

In this talk I will present a recent study
\cite{corge} on supersymmetric contributions to $Z'$ decays, discussing
the impact which they may have on the present searches and mass
exclusion limits.
This investigation updates the pioneering work of
Ref.~\cite{gherghetta} and improves Refs.~\cite{baum,chang} by 
consistently including 
the so-called D-term correction, due to the extra U(1)$'$ group, 
to the sfermion masses.

The U(1)$'$ groups result from the breaking of a rank-6
GUT group ${\rm E}_6\to {\rm SO}(10)\times
{\rm U}(1)'_\psi$, followed by ${\rm SO} (10)\to {\rm SU}(5)
\times {\rm U}(1)'_\chi$.
The heavy neutral bosons associated with ${\rm U}(1)'_\psi$ and 
${\rm U}(1)'_\chi$ are thus 
named $Z'_\psi$ and $Z'_\chi$,
whereas a generic $Z'$ boson is a combination of $Z'_\psi$ and $Z'_\chi$,
with a mixing angle $\theta$:
\begin{equation}
Z'(\theta)=Z'_\psi\cos\theta-Z'_\chi\sin\theta.
\end{equation}
The $Z'$ bosons and the $\theta$ values which will be
investigated are listed in Table~\ref{tabmod}.
The $Z'_\eta$ model comes from the direct breaking of the GUT group in
the SM, i.e. ${\rm E}_6\to {\rm SM} 
\times {\rm U}(1)'_\eta$;
the $Z'_{\rm S}$ is predicted in the secluded model, wherein the SM is extended by means of
a singlet field $S$; the $Z'_{\rm N}$ is instead
equivalent to the $Z'_\chi$,
but with the `unconventional' assignment of SM, MSSM 
and exotic fields 
in the SU(5) representations \cite{nardi}.
\begin{table}
  \caption{$Z'$ bosons in U(1)$'$ gauge models\label{tabmod}}\vspace{0.2cm}
\begin{center}
  \begin{tabular}{|c|c|}
\hline
    Model & $\theta$ \\
\hline
$Z'_\psi$ & 0\\
\hline
$Z'_\chi$ & $-\pi/2$\\
\hline
$Z'_\eta$ & $\arccos\sqrt{5/8}$\\ 
\hline
$Z'_{\rm S}$ & $\arctan(\sqrt{15}/9)-\pi/2$\\
\hline
$Z'_{\rm I}$ & $\arccos\sqrt{5/8}-\pi/2$\\
\hline
$Z'_{\rm N}$ & $\arctan\sqrt{15}-\pi/2$\\
\hline
  \end{tabular}
\end{center}
\end{table}
Because of the $Z'$, the MSSM spectrum gets somewhat modified:
two extra neutralinos are present, for a total
of six neutralinos ($\tilde\chi^0_1,\dots\,\tilde\chi^0_6$), and a novel neutral scalar Higgs,
named $H'$ in \cite{corge}, must be included
to give mass to the $Z'$.
Although, for the sake of consistency, such new particles are
to be taken into account in any phenomenological
analysis, for reasonable choices of the points in
the parameter space, they are too heavy to
contribute to the $Z'$ width and can be safely neglected \cite{corge}.

In the MSSM, the $Z'$ can decay into slepton, squark, chargino, neutralino 
and Higgs pairs,
as well as into final 
states with Higgs bosons associated with a $W$ or a $Z$.
Among these modes, the most interesting ones are those leading to leptonic
final states via supersymmetric particles.
For example,
decays into charged sleptons 
$Z'\to\tilde\ell^+\tilde\ell^-$, with the sleptons decaying 
according to $\tilde\ell^\pm\to \ell^\pm\tilde\chi^0_1$, 
or chargino modes like 
$Z'\to\tilde\chi_2^+\tilde\chi_2^-$, 
followed by $\tilde\chi_2^\pm\to \ell^\pm\tilde\chi_1^0$,
yield final states with two charged leptons and missing energy (neutralinos).
Four leptons and missing energy are instead given by the decay chain
$Z'\to\tilde\chi_2^0\tilde\chi_2^0$, with
subsequent $\tilde\chi_2^0\to \ell^\pm\tilde\ell^\mp$ and 
$\tilde\ell^\pm\to\ell^\pm\tilde\chi_1^0$.
Sneutrino-pair production, i.e.
$Z'\to\tilde\nu_2\tilde\nu_2^*$, followed by  $\tilde\nu_2\to  \tilde\chi^0_2\nu$
and $\tilde\chi_2^0\to \ell^+\ell^-\tilde\chi^0_1$, 
with an intermediate charged slepton, leads to four
charged leptons and missing energy (neutrinos and neutralinos) as well.

The choice of the point in the parameter space to carry out this analysis
is crucial. In fact, this must be done in such a way 
to obtain a supersymmetric spectrum which has not been excluded yet by the
searches at the LHC. Moreover, the lightest Higgs of the MSSM,
which roughly plays the role of the Standard Model one, has to have a mass
about 125-126 GeV, consistent with the recent discovery of a Higgs-like
particle at the LHC \cite{higgs}.
Finally, for this investigation to be meaningful, 
the branching fraction of the $Z'$ into BSM final states has to be 
large enough to motivate searches of the $Z'$ within supersymmetry
and claims that the $Z'$ has not been discovered yet since its 
decay ratios into electron and muon pairs are substantially lower than what the
Standard Model predicts.

Following \cite{corge}, the following reference point can be chosen:
\begin{eqnarray}
&\ &
\mu =200~{\rm GeV}\ ,\ \tan\beta=20\ ,\nonumber\\
&\ & A_q=A_\ell=500~{\rm GeV}\ ,\nonumber\\
&\ & m^0_{\tilde q}=5~{\rm TeV}\ , M_1=150~{\rm GeV}\ ,\nonumber\\
&\ & M_2=300~{\rm GeV}\ ,\ M^\prime=1~{\rm TeV}.
\label{refpoint}
\end{eqnarray}
In Eq.~(\ref{refpoint}) the standard MSSM
notation is adopted: $\mu$ is the parameter
in the 
Higgs superpotential, $\tan\beta=v_2/v_1$ is the ratio of the vacuum
expectation values of the two MSSM Higgs doublets, 
$A_q$ and $A_\ell$ are the couplings of the Higgs with quarks
and leptons.
Furthermore, $m^0_{\tilde q}$ is the squark mass, assumed
to be the same for all flavours at the $Z'$ scale, before the
addition of the D-term; $M_1$, $M_2$ and $M'$ are the soft masses of
the gauginos $\tilde B$, $\tilde W_3$ and $\tilde B '$.
As in \cite{corge}, 
$m^0_{\tilde\ell}$, the slepton mass before the D-term
contribution, is not fixed, but it is just set to
the same value for all selectrons, smuons, staus and sneutrinos, 
and tuned to maximize the $Z'\to \tilde\ell^+\tilde\ell^-$
branching fraction in each model.
As for the couplings, the U(1)$'$ and U(1) ones
are taken to be proportional 
through $g'=\sqrt{5/3}\ g_1$, as occurs in GUTs.
In the Sequential Standard Model, 
the coupling of the $Z'_{\rm SSM}$ to the fermions
is the same as in the SM, i.e. $g_{\rm{SSM}}=g_2/(2\cos\theta_W)$,
where $g_2$ is the SU(2) coupling and $\theta_W$ the Weinberg angle.
The partial widths of the $Z'$ into all decay channels can be found in
\cite{gherghetta}. As for the $Z'_{\rm SSM}$, one assumes that its
couplings to the supersymmetric particles are the same as those of the
Standard Model $Z$ boson.

In the reference point (\ref{refpoint}), as presented in \cite{corge},
the total branching ratio into BSM channels can be up to about 60\%
($Z'_{\rm{SSM}}$), 40\% ($Z'_\psi$), 30\% ($Z'_\eta$ and 
$Z'_{\rm N}$), 20\% ($Z'_{\rm S}$) and 15\% ($Z'_{\rm I}$).
Decays into charged-slepton pairs account for few percent,
whereas a major role is played by final states with chargino and
neutralino pairs, which can be up to 20\% and 30\% of the total
$Z'$ width. Also, in the model $Z'_\eta$,  one has a non-negligible
contribution of sneutrino production, which can be of the order of
10\% \cite{corge}.

To have an idea of the number of events with
supersymmetric $Z'$ decays at the LHC, in 
Table~\ref{number} I present the expected rate of
supersymmetric cascades ($N_{\rm casc}$), 
i.e. the sum of events with 
neutralinos, charginos and sleptons, 
as well as the pure charged-slepton rates ($N_{\rm slep}$), at the LHC
for an integrated luminosity ${\cal L}=20~{\rm fb}^{-1}$
and $\sqrt{s}=8$~TeV. 
All parameters are fixed to
the reference point (\ref{refpoint}); 
the $Z'$ mass is set to either 1.5 or
2 TeV, with $m^0_{\tilde\ell}$ fixed to 
enhance the $Z'\to\tilde\ell^+\tilde\ell^-$ decay
rate \cite{corge}. 
The numbers in Table~\ref{number} are obtained in the
narrow-width approximation, with the $pp\to Z'$ cross section
computed using the leading-order CTEQ6L parton 
distribution functions. The $Z'_\chi$ model
is not taken into account, since, using the parametrization
(\ref{refpoint}), it does not lead to a physical sfermion
spectrum \cite{corge}.
The cascade events can be  
${\cal O}(10^3)$ and the charged sleptons up
a few dozens; the highest rate of 
of supersymmetric particles occurs in the SSM, which enhances the
$Z'$ couplings and thus the production and decay rates,
but even the U(1)$'$ models yield meaningful sparticle production.
\begin{table}
  \caption{Number of supersymmetric cascade events
 and charged sleptons at the LHC
for an integrated luminosity of 20 fb$^{-1}$ and a 
centre-of-mass energy
of 8 TeV. The $Z'$ mass is quoted in TeV.}\vspace{0.2cm}
\begin{center}  
\begin{tabular}{|c|c|c|c|c|c|}
\hline
Model & $m_{Z'}$ & N$_{\rm casc}$ & N$_{\rm slep}$ \\ 
\hline
$Z'_\eta$ & 1.5  & 523  & -- \\
\hline
$Z'_\eta$ & 2 &  55 & -- \\
\hline
$Z'_\psi$ & 1.5 &  599 & 36 \\
\hline
$Z'_\psi$ & 2 & 73 & 4 \\
\hline
$Z'_{\rm N}$ & 1.5 &  400 & 17 \\
\hline
$Z'_{\rm N}$ & 2 &  70 & 3 \\
\hline
$Z'_{\rm I}$ & 1.5  &  317 & -- \\
\hline
$Z'_{\rm I}$ & 2  & 50 & -- \\
\hline
$Z'_{\rm S}$ & 1.5  & 30 & -- \\
\hline
$Z'_{\rm S}$ & 2  & 46  & -- \\
\hline
$Z'_{\rm SSM}$ & 1.5  &  2968  & 95 \\
\hline
$Z'_{\rm SSM}$ & 2  & 462 & 14 \\
\hline
  \end{tabular}
\end{center}
  \label{number}
\end{table}\par
\begin{figure}
\centering
\includegraphics[width=5.4cm,clip]{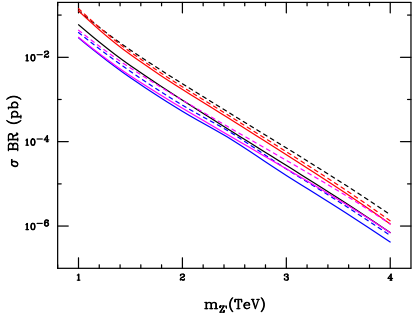}
\caption{Product of the cross section ($\sigma$) times the branching ratio (BR)
into $e^+e^-$ and $\mu^+\mu^-$ pairs for 
$Z'$ production in $pp$ collisions at $\sqrt{s}=8$~TeV,
according to the models $Z'_{SSM}$ (black online), $Z'_\eta$ (blue),
$Z'_{\rm I}$ (red) and 
$Z'_\psi$ (magenta). The solid lines account for BSM modes, the dashes
just rely on SM channels.}
\label{sbr}
\end{figure}

Before concluding, as in \cite{blois},
since the experimental analyses search for high-mass
dielectron and dimuon pairs, I present the results 
in terms of the product of the $Z'$ production
cross section ($\sigma$) at the LHC ($\sqrt{s}=8$~TeV)
times the branching ratio (BR) into $e^+e^-$ and $\mu^+\mu^-$
pairs, with and without accounting for 
supersymmetric modes.
In fact, the experimental exclusion limits are obtained by comparing
data and theoretical predictions for the product $\sigma $~BR.
Fig.~\ref{sbr} shows this product 
varying the $Z'$ mass in the range 1 TeV$<m_{Z'}<$ 4 TeV,
with only SM decays (dashes) and accounting for possible
supersymmetric contributions (solid lines). In the BSM case,
the parameters are fixed to the reference point (\ref{refpoint}),
with the slepton mass $m^0_{\tilde\ell}$ set as explained above
\cite{corge}.
One can thus note that, when including the 
BSM decay modes, the suppression of $\sigma$~BR is about 60\% for 
the $Z'_{\rm SSM}$, 40\% for the $Z'_\psi$ model, 30\% for the $Z'_\eta$ and
13\% for the $Z'_{\rm I}$.
Such results point out a remarkable impact of the inclusion of the
supersymmetric contributions to $Z'$ decays and 
that a novel analysis, taking into account
such modes, may be worthwhile to be pursued.
\begin{figure}[ht!]
\centering
\includegraphics[width=5.4cm,clip]{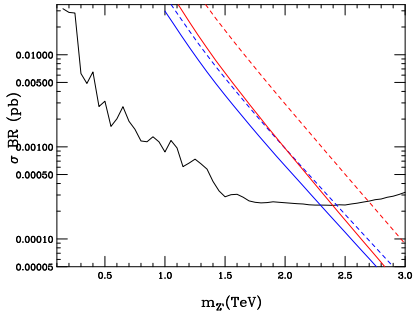}
\caption{Comparison of the product $\sigma(pp\to Z')~{\rm BR} 
(Z'\to \ell^+\ell^-)$, as measured by the ATLAS Collaboration (black solid line)
with the corresponding theory predictions including supersymmetric decays (solid lines)
and only SM modes (dashes) for the $Z'_{\rm SSM}$ (red) and $Z'_\psi$ (blue) models.}
\label{sigbatlas}
\end{figure}
\par
\begin{figure}[ht!]
\centering
\includegraphics[width=5.4cm,clip]{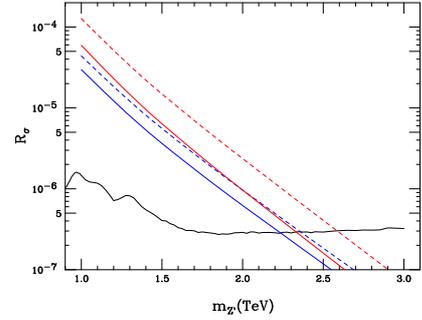}
\caption{As in Fig.~\ref{sigbatlas}, but comparing the theory predictions with the
CMS measurement of the ratio $R_\sigma=[\sigma(pp\to Z')~{\rm BR} 
(Z'\to \ell^+\ell^-)]/[\sigma(pp\to Z)~{\rm BR} 
(Z\to \ell^+\ell^-)]$.}
\label{sigbcms}
\end{figure}

 Finally, I compare the ATLAS 
(Fig.~\ref{sigbatlas}) and CMS (Fig.~\ref{sigbcms}) data on $\sigma~{\rm BR}$ with
our predictions with and without supersymmetric $Z'$ decays. As for CMS
in the plotted
ratio $R_\sigma$ the high-mass dilepton data are normalized
to the $\sigma~{\rm BR}$ product for Standard Model $Z$ production.
From such a comparison, one can learn that, in the reference point, the inclusion of supersymmetric
contributions to $Z'$ decays can have an impact of about 200-300 GeV on the
mass exclusion limits.

In summary, I discussed possible supersymmetric signals in the decays of new
neutral vector bosons $Z'$, predicted in GUT-inspired U(1)$'$ groups and in the
Sequential Standard Model, paying special attention to decay modes into sleptons, charginos
and neutralinos, leading to final states with charged leptons and missing energy.
For a representative point of the phase space, it was found that, especially in the
SSM, the supersymmetric decays can be competitive with the standard ones, up to the point of 
lowering by 200-300 GeV the current limits on the $Z'$ mass. However,
in order to make a final statement on this issue, a novel experimental analysis including
supersymmetric channels is compelling.
\end{document}